\documentclass[12pt]{article}

\usepackage{scicite}

\usepackage{times}
\usepackage{pdfpages}
\usepackage{pgffor}

\usepackage{hyperref}

\newenvironment{sciabstract}{%
\begin{quote} \bf}
{\end{quote}}

\newcounter{lastnote}

\usepackage{graphicx}%
\usepackage{dcolumn}%
\usepackage{bm}%

\usepackage{xcolor}
\usepackage[normalem]{ulem}
\usepackage{mhchem}

\newcommand{\twocol}{12cm}

\title{Machine Learning Unifies the 
Modelling \\of Materials and Molecules}

\author{Albert P. Bart\'ok,$^{1}$ 
Sandip De,$^{2,3}$ 
Carl Poelking,$^{4}$\\
Noam Bernstein,$^{5}$
James Kermode,$^{6}$
G{\'a}bor Cs{\'a}nyi,$^{7}$
Michele Ceriotti$^{3,\ast}$\\
\\
\normalsize{$^{1}$Scientific Computing Department, Science and Technology Facilities Council,}\\\normalsize{ Rutherford Appleton Laboratory,} \normalsize{Oxfordshire OX11 0QX, United Kingdom}\\
\normalsize{$^{2}$National Center
for Computational Design and Discovery}\\\normalsize{of Novel Materials (MARVEL)}\\
\normalsize{$^{3}$Laboratory of Computational Science and Modelling,}\\\normalsize{Institute of Materials, EPFL, Lausanne, Switzerland}\\
\normalsize{$^{4}$Department of Chemistry,}\\\normalsize{University of Cambridge, Cambridge CB2 1EW, United Kingdom}\\
\normalsize{$^{5}$Center for Materials Physics and Technology,}\\\normalsize{ U.S. Naval Research Laboratory, Washington, DC 20375, USA}\\
\normalsize{$^{6}$Warwick Centre for Predictive Modelling,}\\\normalsize{ School of Engineering, University of Warwick, Coventry CV4 7AL, United Kingdom}\\
\normalsize{$^{7}$Engineering Laboratory, University of Cambridge}
\\
\normalsize{$^\ast$To whom correspondence should be addressed; E-mail:  michele.ceriotti@epfl.ch.}
}

\date{}
\begin{document}

\baselineskip24pt

\maketitle 
\newpage
\begin{sciabstract}
Determining the stability of molecules and condensed phases is the cornerstone of
atomistic modelling, underpinning our understanding of chemical and materials
properties and transformations.
Here we show that a machine-learning model, 
based on a local description of chemical environments and Bayesian statistical 
learning, provides a unified framework to predict atomic-scale properties.
It captures the quantum mechanical effects governing the complex 
surface reconstructions of silicon, predicts the stability of
different classes of molecules with chemical accuracy, 
and distinguishes active and inactive protein ligands with more than 99\% reliability.  The universality and the
systematic nature of our framework provides new insight into the potential 
energy surface of materials and molecules.
\end{sciabstract}

{\bf\noindent RESEARCH SUMMARY: Statistical learning based on a local representation of atomic
structures provides a universal model of chemical stability}

\section{Introduction}

Calculating the energies of molecules and of condensed-phase structures is fundamental to predicting the behavior of 
matter at the atomic scale, and a formidable challenge. Reliably assessing the relative stability
of different compounds, and of different phases of the same material, 
requires the evaluation of the energy of a given three-dimensional assembly of
atoms with an accuracy comparable with the thermal energy ($\sim$0.5~kcal/mol at 
room temperature), which is a small fraction of the energy of a chemical bond (up to $\sim$230~kcal/mol for the \ce{N2} molecule). 

Quantum mechanics is a universal framework that can deliver this level of accuracy. By solving the Schr\"odinger equation, the electronic structure of materials and molecules can
in principle be computed, and from it all 
ground-state properties and excitations follow. 
The prohibitive computational cost of exact solutions 
at the level of electronic-structure theory lead to the development of many approximate techniques that address different classes of systems. Coupled-cluster theory (CC)\cite{SzaboOstlund} for molecules, and density-functional theory (DFT)\cite{RichardMartinDFT,DFT-HK,DFT-KS}, for the condensed phase have been particularly successful and can typically deliver the levels of accuracy required to address a plethora of important scientific questions. 
The computational cost of these electronic structure models is nevertheless still significant, limiting their routine application in practice to dozens of atoms in the case of CC and hundreds in the case of DFT. 

To go further, explicit electronic structure calculation
has to be avoided, and we have to predict the energy corresponding to an atomic configuration directly. While such empirical potential methods (force fields) are indeed much less expensive, their predictions to date have been qualitative at best. 
Moreover, the number of distinct approaches have rapidly multiplied -- in the struggle for accuracy at low cost, generality is invariably sacrificed. Recently,  machine-learning approaches have started to be applied
to designing interatomic potentials  that {\em interpolate}
electronic-structure data as opposed to using parametric functional forms tuned to match experimental or calculated observables.
While there have been several
hints that this approach can achieve the
accuracy of DFT at a fraction of the
cost\cite{behl-parr07prl,bart+10prl,snyd+12prl,mont+13njp,fabe+16prl,shap16mms,ANI1}, little effort has been put into recovering the generality of quantum mechanics: atomic and molecular descriptors, and 
learning strategies have been optimized
for different classes of problems, and in particular efforts for materials and for chemistry have been rather disconnected. 
Here we show that the combination of Gaussian process regression\cite{RasmussenBook} with a local 
descriptor of atomic neighbor environments that is 
general and systematic can re-unite the 
modelling of hard matter and molecules, consistently achieving 
predictive accuracy. 
{
This lays the foundations for a 
universal reactive force field that can recover the accuracy of the Schr\"odinger equation at negligible cost and -- thanks to the locality of the model -- leads to an intuitive understanding of the stability and the interactions between molecules. 
By showing that we can accurately classify active and inactive protein ligands we provide evidence that this framework can be extended to capture more complex, non-local properties as  well.}

\begin{figure*}[bth]
\centering\includegraphics[width=\twocol]{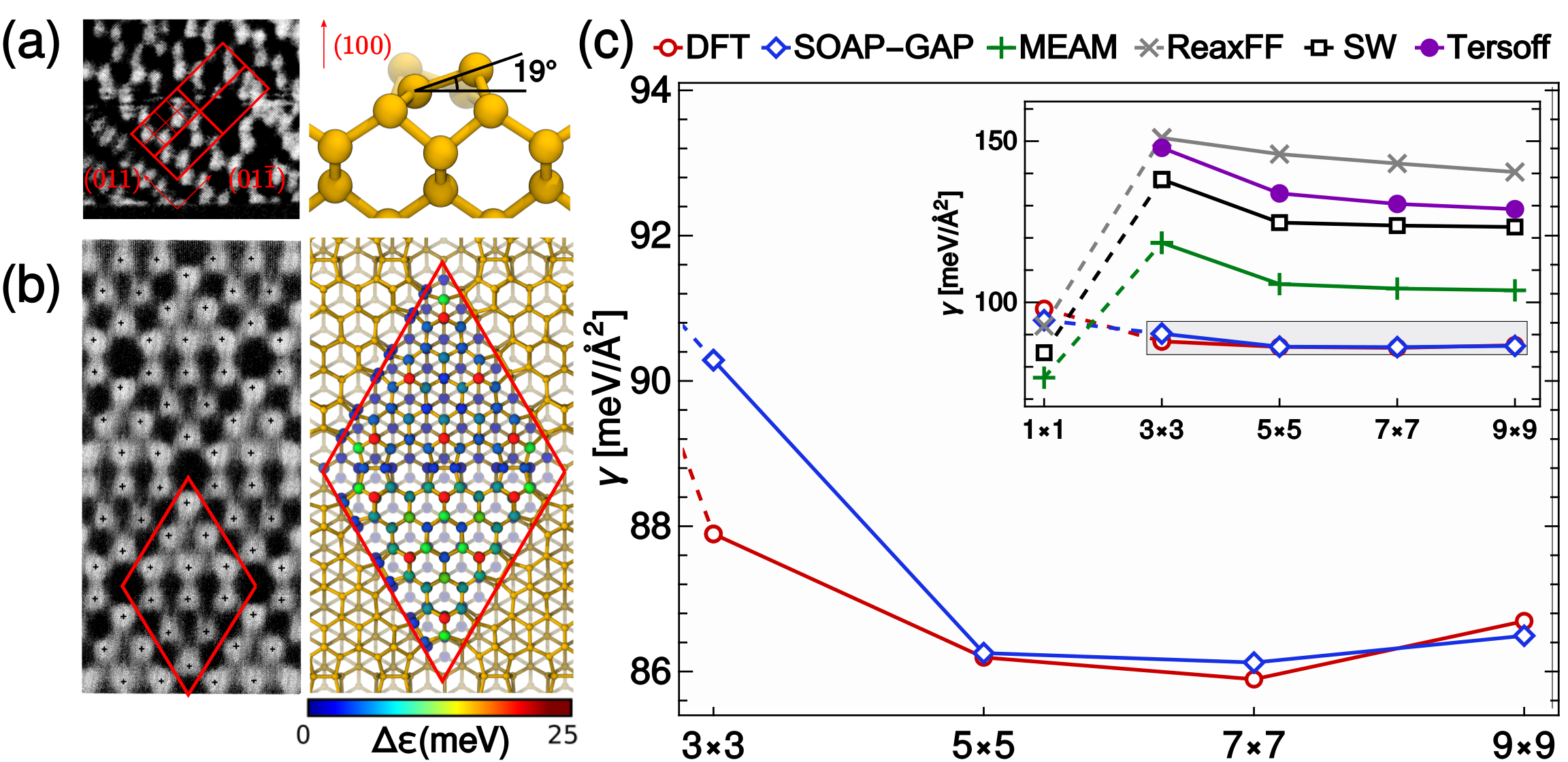}
\caption{(a) The tilt angle of dimers on the reconstructed Si(100) surface (STM image left\cite{wolk92prl}, SOAP-GAP relaxed structure right) are the result of a Jahn-Teller distortion, predicted to be about 19$^\circ$ by DFT and SOAP-GAP. Empirical force fields show no tilt.
(b) The Si(111)-7$\times$7 
reconstruction
is an iconic example of the complex
structures that can emerge from the
interplay of different quantum mechanical
effects (left: STM image\cite{binn+83prl}, right: SOAP-GAP relaxed structure colored 
by predicted local energy error {when using a training set without adatoms}); (c) reproducing this delicate
balance and predicting that the $7\times 7$ 
is the ground-state structure is one of 
the historical successes of DFT: a 
SOAP-based machine-learning model is the
only one that can describe this ordering, while widely used force fields 
incorrectly predict the un-reconstructed
surface (dashed lines) to have a lower energy state.
}
\label{fig:silicon}
\end{figure*}

\section{Results}

\subsection{The reconstructions of silicon surfaces}

 Due to its early technological relevance to the semiconductor industry and simple bulk structure,
Si has traditionally been one of the archetypical tests for 
new computational approaches to materials modelling~\cite{car-parr85prl,Rinke2009,Williamson2002,behl-parr07prl,behl+08prl,bart+10prl}. In spite of 
the fact that its bulk properties can be captured reasonably
well by simple empirical potentials, its surfaces display
remarkably complex reconstructions, whose stability
is governed by a subtle balance of elastic properties and  quantum mechanical 
effects, such as the Jahn-Teller distortion that determines
a tilt of dimers on Si(100). The determination of the dimer-adatom-stacking fault (DAS) 
$7\times 7$ reconstruction of Si(111) as the most stable 
among several similar structures was the culmination of 
a concerted effort of experiment and modelling
including early scanning tunnelling microscopy~\cite{binn+83prl}, and was also a triumph for DFT~\cite{brom+92prl}.

As shown in Figure~\ref{fig:silicon}, 
empirical potentials incorrectly predict the 
un-reconstructed $1\times 1$ to be a lower 
energy configuration, and fail to predict the $7\times7$ as the lowest energy structure even from among the DAS reconstructions. Up to now, the only models that
could capture these effects included electronic 
structure information, at least on the tight binding level (or its 
approximation as a bond order potential).
We trained a SOAP-GAP model on a database of configurations from short {\em ab initio} molecular dynamics trajectories of small unit cells {(including the $3\times 3$ reconstruction, but not those with larger unit cells; for details, see SI)}. 
This model correctly describes
a broad array of standard bulk and defected material properties within a wide range of pressures and temperatures, {as well as properties that depend on transition state energetics such as the generalized stacking fault surfaces shown in the SI}. A striking illustration of the power of this model
is the quantitative description of both the tilt of the (100) dimers and the 
ordering of the (111) reconstructions -- without explicitly considering the quantum mechanical
electron density. 

Nevertheless, even this model is based on a training dataset which is a result of ad hoc (if well informed) choices. The Bayesian GPR framework tells us how to improve the model. The predicted error $\sigma^*$, shown as the color scale in 
Fig.~\ref{fig:silicon}b, can be used to identify new configurations likely to be usefully added to the training set. The adatoms on the surface have the highest error, {and once we included small surface unit cells with adatoms, the ML model came much closer to its target}.

\begin{figure*}[hbtp]
\centering\includegraphics[width=\textwidth]{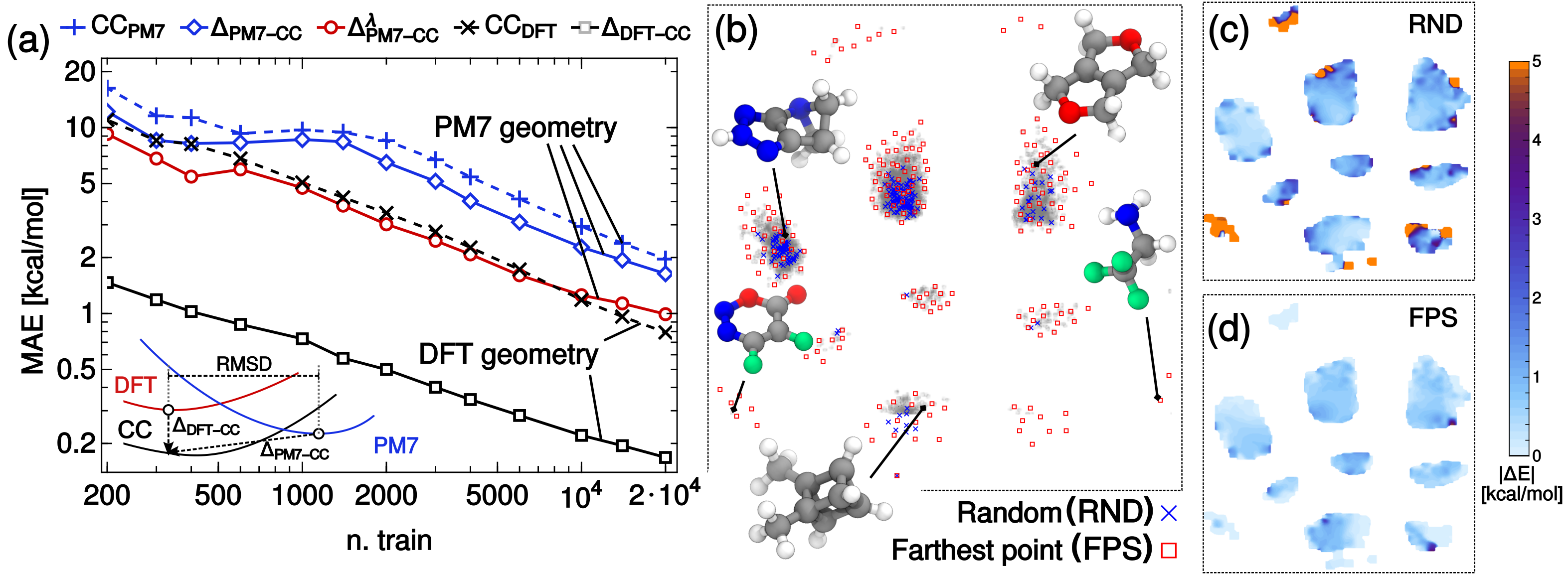}
\caption{%
(a) Learning curves for the coupled-cluster (CC) atomization energy of molecules in the
GDB9 dataset, using the average-kernel SOAP with a cutoff of 3~\AA{}.
Black lines correspond to using DFT geometries to predict CC energies for the DFT-optimized geometry.
Using  the DFT energies as a baseline and learning
$\Delta_\text{DFT-CC}=E_\text{CC}-E_\text{DFT}$ lead to a five-fold 
reduction of the test error compared to learning CC energies directly 
as the target property (CC$_\text{DFT}$). 
The other curves correspond to using  PM7-optimized 
geometries as the input to the prediction of CC energies 
of the DFT geometries. There is little improvement when 
learning the energy correction ($\Delta_\text{PM7-CC}$) 
compared to direct training on the CC energies (CC$_\text{PM7}$). 
Using information on the structural 
discrepancy between PM7 and DFT geometries in the \emph{training} set, 
however, brings the prediction error down to 1~kcal/mol MAE ($\Delta^\lambda_\text{PM7-CC}$, with $\lambda=0.25$\AA).
(b) A sketch-map representation of the GDB9 (each gray point corresponding to one structure) highlights the importance of 
selecting training configurations to uniformly cover configuration space. The average prediction error for different portions of the map is markedly different when using a random selection (c) and farthest point sampling (d). The latter is much better behaved in the peripheral, poorly-populated regions. 
}
\label{fig:gdb9}
\end{figure*}

\subsection{Coupled-cluster energies for 130k Molecules }

Molecular properties exhibit distinctly different 
challenges than bulk materials, from 
the combinatorial number of stable configurations, 
to the presence of collective quantum mechanical
and electrostatic phenomena such as aromaticity, 
charge transfer and hydrogen bonding. At the 
same time, many relevant scientific questions 
involve predicting the energetics of stable
conformers, which is a less-complex problem than
obtaining a reactive potential.
Following early indication of success on a small 
dataset~\cite{mont+13njp,de+16pccp},
here we start our investigation using the GDB9 dataset
that contains about 134,000 small organic molecules 
whose geometries have been optimized at the level
of DFT, and that 
has been used in many of the 
pioneering studies of machine learning
for molecules~\cite{rama+14sd,rama+15jctc}. Accurate
models have been reported, however, only when
predicting DFT-energies using as inputs geometries
that have already been optimized at the DFT level
 --
which makes the exercise
insightful~\cite{schu+17ncomm} but does not constitute an alternative to doing the DFT calculation.  %

Figure~\ref{fig:gdb9}a demonstrates that the
GPR framework using the very same SOAP descriptors can be
used to obtain \emph{useful predictions} of the chemical
energy of a molecule (the atomization energy) on this %
heterogeneous chemical dataset. 
DFT methods give very good equilibrium geometries,
and are often used as a stepping stone to evaluate energies at the
``gold standard'' level of CC theory (CCSD(T)). They have also been
shown to constitute an excellent baseline reference 
towards higher levels of theory~\cite{rama+15jctc}.
Indeed, a SOAP-GAP model can use 
DFT inputs and only 500 training points 
to predict CCSD(T) atomization energies with an 
error below the symbolic threshold of 1~kcal/mol. 
The error drops to less than 0.2~kcal/mol when training on 
15\%{} of the GDB9.

DFT calculations for the largest molecules in GDB9 can nowadays
be performed in a few hours, which is still impractical if one 
wanted to perform high-throughput molecular screening on millions 
of candidates. 
Instead, we can use the inexpensive semi-empirical PM7 model 
(taking around a second to compute a typical GDB9 molecule) to obtain 
an approximate relaxed geometry, and build a model to bridge the gap 
between {\em geometries and energies}\cite{rama+15jctc}. With a training 
set of ~20,000 structures, the model predicts CCSD(T) energies with 
1~kcal/mol accuracy using only the PM7 geometry and energy as input. 

In order to achieve this
level of accuracy it is however
crucial to use this information
judiciously. 
The quality of PM7 training points,
as quantified by the root-mean square difference (RMSD) $d$ 
between PM7 and DFT geometries, varies significantly across the GDB9.
 In keeping with the Bayesian spirit of the ML framework,
we set the diagonal variance 
$\propto \exp (d^2/\lambda^2)$
corresponding to the prior information that structures with a larger RMSD 
between the two methods may be affected by a larger 
uncertainty. Even though we {\em do not} use RMSD information
on the test set, the effect of down-weighting  information from the training
structures for which PM7 gives inaccurate geometries 
is to reduce the prediction error by more than 40\%{}.

The strategy used to 
select training structures also has a significant
impact on the reliability of the model.
Fig.~\ref{fig:gdb9}b shows a 
sketch-map~\cite{ceri+11pnas} of the structure of the GDB9 dataset
based on the kernel-induced metric, demonstrating
the inhomogeneity of the density of 
configurations. Random selection of reference 
structures leaves large portions of the 
space unrepresented, which results in a very 
heavy-tailed distribution of errors (see SI). We find that
selecting the training set sequentially using a 
greedy algorithm that picks the next farthest 
data point to be included (farthest point sampling, FPS)
gives more uniform sampling of the database, 
 dramatically reducing the fraction of large errors,
especially in the peripheral regions of the dataset (Fig.~\ref{fig:gdb9}c and d),
leading to a more resilient ML model. 
It should be noted that this comes at the price of a small degradation 
of the performance as measured by the commonly used mean absolute error (MAE),
due to the fact that densely populated regions 
do not get any preferential sampling.

In order to test the ``extrapolative power'', or transferability
of the SOAP-GAP framework we then
applied the GDB9-trained model for $\Delta_\text{DFT-CC}$ to the prediction
of the energetics of larger molecules, and
considered $\sim 850$ conformers of the 
dipeptides obtained from two natural amino acids, aspartic
acid and glutamic acid~\cite{ropo+16sd}. Although GDB9 does not
explicitly contain information on the 
relative energies of conformers of the same molecule, 
we could predict the CCSD(T) corrections to the 
DFT atomization energies
with an error of 0.45 kcal/mol, a 100-fold 
reduction compared to the intrinsic error of DFT. 

{It is worth stressing that, 
within the scope of the SOAP-GAP framework, there is considerable room for improvement of the accuracy. Using the same SOAP parameters that we adopted for the GDB9 model for the benchmark task of learning DFT energies using DFT geometries as inputs, we could obtain a mean absolute error of 0.40~kcal/mol in the smaller QM7b dataset~\cite{mont+13njp}.
As discussed in the SI, using an ``alchemical kernel''~\cite{de+16pccp} to include correlations between different species allowed us to further reduce that error to 0.33~kcal/mol. A ``multi-scale'' kernel (a sum of SOAP kernels each with a different radial cutoff parameter) that combines information from different length scales allows one to reach an accuracy of 0.26 kcal/mol (or alternatively, to reach 1~kcal/mol accuracy with fewer than 1000 FPS training points) -- both results being considerably superior to existing methods that have been demonstrated on similar datasets.
{The same multi-scale kernel also improves significantly the performance for GDB9, allowing us to reach 1~kcal/mol with just 5000 reference energies, and as little as 0.18 kcal/mol with 75000 structures.}
Given that SOAP-GAP allows naturally to both predict and learn from derivatives of the potential (i.e. forces), the doors are open for building models that can describe local fluctuations and/or chemical reactivity by extending the training set to non-equilibrium configurations -- as we demonstrated already for the silicon force field here, and previously for other elemental materials. 
}

\begin{figure}[bt]
\centering\includegraphics[width=\twocol]{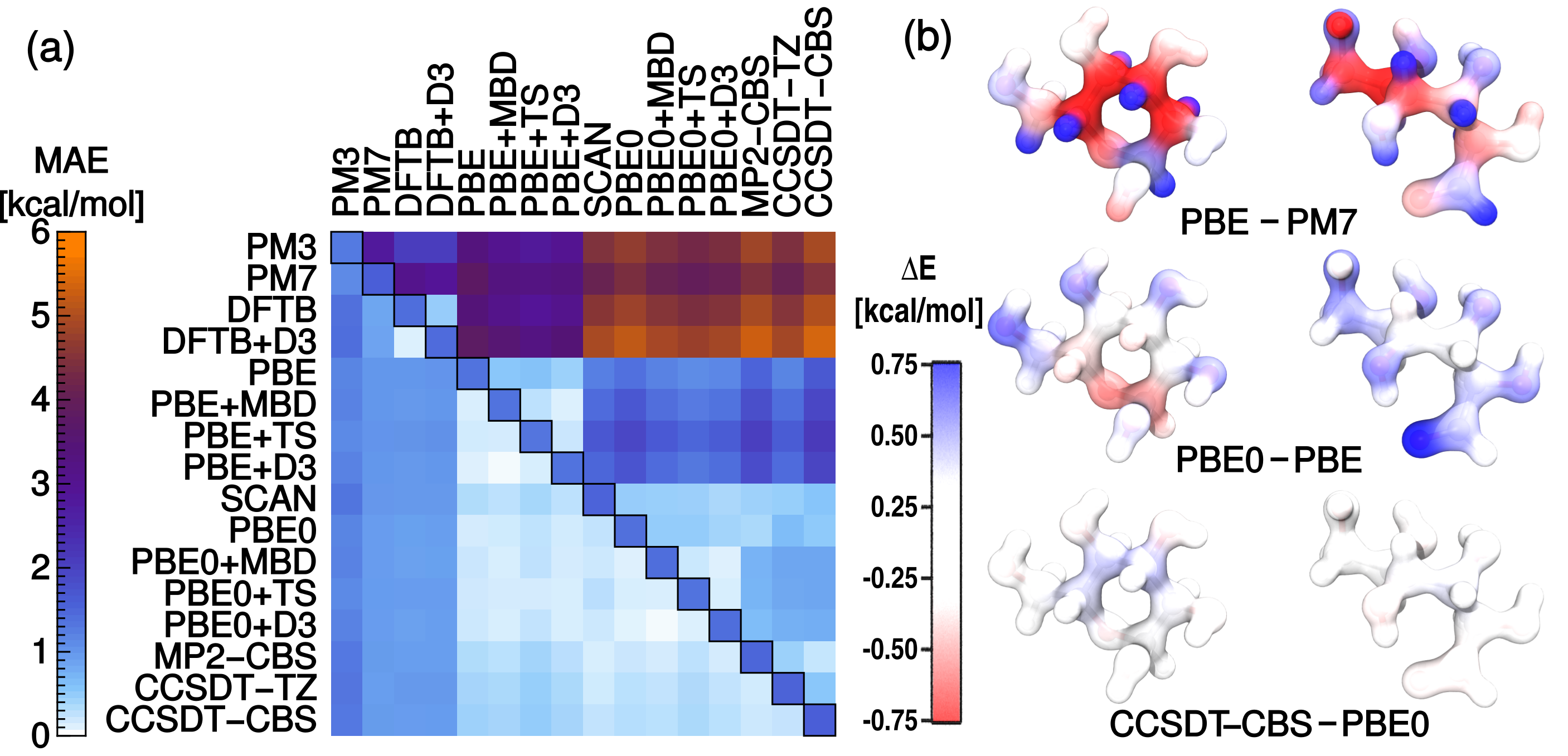}
\caption{ (a) Extensive tests on 208 
conformers of glucose (taking only 20 FPS samples for
training) reveals the potential of a ML approach 
to bridge different levels of quantum chemistry; the 
diagonal of the plot shows the MAE resulting
from direct training on each level of theory;
the upper half shows the intrinsic difference between 
each pairs of models; the lower half
shows the MAE for learning each correction. 
{(b) The energy difference between three pairs of  electronic-structure methods, partitioned in atomic contributions based on a SOAP analysis, and represented as a heat map. The molecule on the left represent the lowest-energy conformer of glucose in the dataset, and the one on the right the highest-energy conformer.}
}
\label{fig:others}
\end{figure}

\subsection{The stability of molecular conformers}

To reduce even further the prediction
error on new molecules, we could include a larger set of 
training points from the GDB9. It is clear 
from the learning curve in Fig.~\ref{fig:gdb9}a
that the ML model is still far from its saturation point. {For the benchmark DFT learning exercise we could attain an error smaller than 0.28 kcal/mol using 100k training points, which can be improved even further using a more complex multi-scale kernel (see SI).} 
An alternative is to train a specialized
model that aims to obtain accurate predictions of the relative energies of a set
of similar molecules. 
As an example of this approach, 
we considered a set of 208 conformers of 
glucose, whose relative stability has been 
recently assessed with a large set of electronic-structure methods~\cite{mari+16jctc}. 
Fig.~\ref{fig:others}a shows that as few as 20 reference
configurations are sufficient to evaluate the 
corrections to semiempirical energies that are needed to reach 1 kcal/mol 
accuracy relative to 
complete-basis-set CCSD(T) energies, or to reach
0.2-0.4 kcal/mol error when using different
flavors of DFT as a baseline.

\begin{figure*}
\centering\includegraphics[width=\twocol]{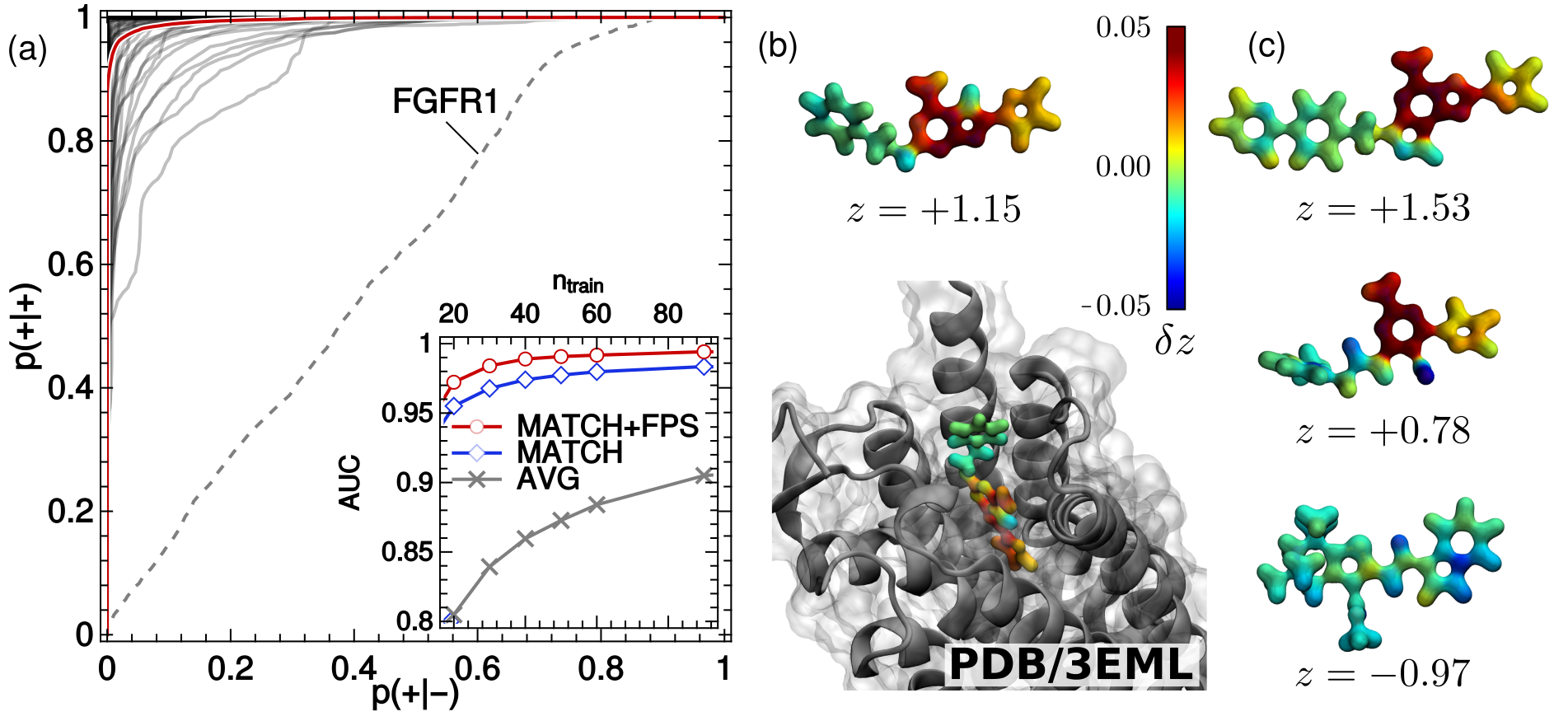}
\caption{
(a) Receiver operating characteristics (ROCs) of binary classifiers based on a SOAP kernel, applied to the
prediction of the binding behavior of ligands and decoys 
taken from the DUD-E, trained on $60$ examples. Each ROC corresponds to one specific protein receptor. The red curve is the average over the individual ROCs. The dashed line 
corresponds to receptor FGFR1, which contains inconsistent
data in the latest version of the DUD-E.
In the inset, the AUC (Area Under the Curve) performance measure as a function of the number of ligands used in the training, for the ``best match''-SOAP kernel (MATCH) and average molecular SOAP kernel (AVG); 
(b-c) 
Visualization of binding moieties for adenosine receptor A2 (AA2AR) as predicted for the crystal ligand (b), as well as two known ligands and one decoy (c). The contribution of an individual atomic environment to the classification is quantified by the contribution $\delta z_i$ in signed distance $z$ to the SVM decision boundary and visualized as a heat map projected on the SOAP neighbor density (images for 
all ligands and all receptors are accessible at \cite{libatoms_dude}). 
Regions with $\delta z > 0$ contain structural patterns expected to promote binding (see color scale and text). The snapshot in (b) indicates the position of the crystal ligand in the receptor pocket as obtained by X-ray crystallography~\cite{jaakola_2.6_2008}.}
\label{fig:dude}
\end{figure*}

\subsection{Receptor ligand binding}

The accurate prediction of molecular energies opens up the possibility
of computing a vast array of more complex thermodynamic properties,
using the SOAP-GAP model as the underlying energy engine in molecular dynamics simulation.
However, the generality of the SOAP kernel for describing chemical environments
also allows directly attacking different classes of scientific
questions -- e.g. sidestepping not only the evaluation
of electronic structure, but also the cost of 
demanding free-energy calculations, making instead a direct connection
to experimental observations. 
As a demonstration of the potential of this approach, 
we  investigated the problem of ligand-receptor binding. 
We used data from the DUD-E (Directory of Useful 
Decoys, Enhanced)~\cite{mysinger_directory_2012}, a highly curated set of ligand-receptor 
pairs taken from the ChEMBL database, enriched with 
property-matched decoys~\cite{lagarde_benchmarking_2015}. These decoys resemble the 
individual ligands in terms of atomic composition, 
molecular weight, and physicochemical properties, but are 
structurally distinct in that they do not bind to the protein receptor.

We trained a Kernel-Support-Vector-Machine (Kernel-SVM)\cite{scho+98nc,scholkopf_learning_2002} for each of the 102 receptors listed in
the DUD-E, to predict whether or not each candidate molecule
binds to the corresponding polypeptide.
We used an equal but varying number $n_\textrm{train}$ of ligands and decoys (up to 120)
for each receptor, using the SOAP kernel as 
before to represent the similarity between atomic environments. Here 
however we chose the matrix $\mathbf{P}$ in eq. (\ref{eq:mol2env}) corresponding to
an optimal permutation matching (``MATCH''-SOAP) rather than a uniform average~\cite{de+16pccp}.
Predictions are collected over the remaining compounds and the results are averaged over different subsets used for training.

The receiver-operating characteristic (ROC), shown in Fig.~\ref{fig:dude}, describes the trade-off between the rate of true positives $p(+|+)$ versus false positives $p(+|-)$ as the decision threshold of the SVM is varied. The area under the ROC curve (AUC) is a widely used performance measure of binary classifiers, in a loose sense giving the fraction of correctly classified items. 
A SOAP-based SVM trained on just 20 examples can predict receptor ligand binding with a typical accuracy of 95\%, which goes up to 98\% when 60 training examples are used, and 99\% when using a FPS training set selection strategy -- 
significantly surpassing the present state-of-the-art\cite{skoda_benchmarking_2016,lee_predicting_2016,wallach_deep_2015}.
The model is so reliable that its failures are highly suggestive of inconsistencies in the underlying data. The dashed line in Fig.~\ref{fig:dude}a corresponds to receptor FGFR1 and shows no predictive capability. Further investigation uncovered data corruption in the DUD-E dataset, with identical ligands labelled both as active and inactive. Using an earlier version of the database~\cite{huang_benchmarking_2006} shows no such anomaly, giving an AUC of 0.99 for FGFR1.

\section{Discussion}

Machine learning is often regarded -- and criticized -- as the quintessentially
na\"ive inductive approach to science. In many cases, however, one
can extract some intuition and insight from a critical look at the behavior of a 
machine-learning model. 

Fitting the difference between  
levels of electronic structure theory 
gives an indication of how smooth and localized,
and therefore easy for SOAP-GAP to learn, are the corrections
that are added by increasingly expensive methods. 
For instance, hybrid DFT methods are considerably 
more demanding than plain ``generalized-gradient approximation''
DFT, and indeed show a considerably smaller baseline variance
to high-end quantum chemistry methods. However, the 
error of the corresponding SOAP-GAP model is almost
the same for the two classes of DFT, which indicates that 
exact-exchange corrections to DFT are particularly short ranged, and
therefore easy to learn with local kernel methods.
{Thanks to the additive nature of the average-kernel SOAP kernel, it is also possible to decompose the energy difference between methods into atom-centered contributions (Fig.~\ref{fig:others}b). 
The discrepancy between DFT and semiempirical methods appears to involve large terms with opposite sign (positive for carbon atoms, negative for aliphatic hydrogens), that partially cancel out. Exact exchange plays an important role in determining the energetics of the ring and open chain forms~\cite{mari+16jctc}, and indeed the discrepancy between PBE and PBE0 is localized mostly on the aldehyde/hemiacetal group, as well as, to a lesser extent, on the H-bonded O atoms.
The smaller corrections between CC methods and hybrid functionals show less evident patterns,
as one would expect when the corrections involve correlation energy. 
}
Long-range non-additive components to the energy are expected for any system with
electrostatic interactions { -- and could be treated, for instance, by machine-learning the local charges and dielectric response terms~\cite{artr+11prb}, and then feeding them into established models of electrostatics and dispersion. 
}
However for elemental materials and the
small molecules we consider here an additive
energy model can be improved simply by increasing the
kernel range, $r_c$. 
Looking at the dependence
of the learning curves on the cutoff for the GDB9 (see SI),
we can observe the trade-off between the completeness 
of the representation and its extrapolative power~\cite{huan-vonl16jcp}.
For small training set sizes, a very
short cutoff of 2~\AA{} and the averaged
molecular kernel give the best
performance, but then saturates at about 2~kcal/mol.
Longer cutoffs
give initially worse performance,  because
the input space is larger, but
the learning rate deteriorates more slowly;
at 20,000 training structures,
$r_c=3$~\AA{} yields the best performance. 
Given that the SOAP kernel gives
a complete description~\cite{bart+13prb} of each 
environment up to $r_c$, we can
infer from these observations the 
relationship between the length and energy scales of 
physical interactions (see SI). 
For a DFT model, considering interactions
up to 2~\AA{} is optimal if
one is content to capture physical 
interactions with an energy scale
of the order of 2.5~kcal/mol.
When learning corrections to electron correlation,
$\Delta_\text{DFT-CC}$, most of the
short-range information is already
included in the DFT baseline, and so
length scales up to and above 3~\AA{}
 become relevant already for
 $n_\text{train}<20,000$, allowing an accuracy 
 of less that 0.2~kcal/mol to be reached.

In contrast, the case of ligand binding predictions poses a significant challenge to an
additive energy model already at the small molecule scale.
Ligand binding is typically mediated by electro-negative/positive or 
polarizable groups located in ``strategic'' locations within the ligand 
molecule, which additionally must satisfy a set of steric constraints 
in order to fit into the binding pocket of the receptor. 
Capturing these spatial correlations of the molecular structure is a prerequisite to accurately predicting whether or not a given molecule binds to a receptor. This is demonstrated by the unsatisfactory performance of
a classifier based on an averaged combination
of atomic SOAP kernels (see Fig.~\ref{fig:dude}b).
By combining the atomic SOAP kernels using an ``environment matching'' procedure,
one can introduce a degree of 
non-locality -- because now environments in the
two molecules must be matched pairwise, rather
than in an averaged sense. 
Thus, the relative performance of different kernel combination strategies
give a sense of whether the global property 
of a molecule can result from averages 
over different parts of the system, or whether
a very particular spatial distribution of 
molecular features is at play.

A striking demonstration of inferring structure-property relations from a
ML model is given in Fig.~\ref{fig:dude}b-c, 
where the SOAP classifier is used to identify binding moieties 
(``warheads'') for each of the receptors. 
To this end, we formally project the SVM decision function $z$ 
onto individual atoms of a test compound associating 
to each a ``binding score'' (see SI).
Red and yellow regions of the isosurface plots denote moieties that are 
expected to promote binding. For decoys, no consistent 
patterns are resolved. The identified warheads 
are largely conserved across the set of ligands -- 
in fact, by investigating the position of the crystal 
ligand inside the binding pocket of the adenosine 
receptor A2 (b), we can confirm that a positive binding 
field is indeed assigned to those molecular fragments that 
localize in the pocket of the receptor. Scanning through the 
large set of ligands in the dataset (see SI), it is also clear that the six-membered ring and its amine group, fused with the adjacent five-membered ring, are the most prominent among the actives.
Finally, note that regions of the active ligands colored in blue (as in Fig.~\ref{fig:dude}c) could serve as target locations for lead optimization, e.g., to improve receptor affinity and selectivity.

The consistent success of the SOAP-GAP framework across materials, 
molecules and biological systems shows that it is possible to sidestep the explicit 
electronic structure and free energy calculation 
and determine the direct relation between molecular
geometry and stability. This already enables useful predictions to be made 
in many problems, and there is a lot of scope for further 
development -- e.g. by using a deep-learning
approach, by developing multi-scale kernels to treat long 
range interactions, using 
active learning strategies\cite{li+15prl}, or by fine tuning the assumed
correlations between the contributions of different chemical elements, as discussed in the SI.
{We believe that the exceptional performance of the SOAP-GAP framework we demonstrated stems from its general, mathematically rigorous approach to the problem of representing local chemical environments. Building on this local representation allowed us to capture even more complex, non-local properties. }

\begin{figure}
\centering\includegraphics[width=7cm]{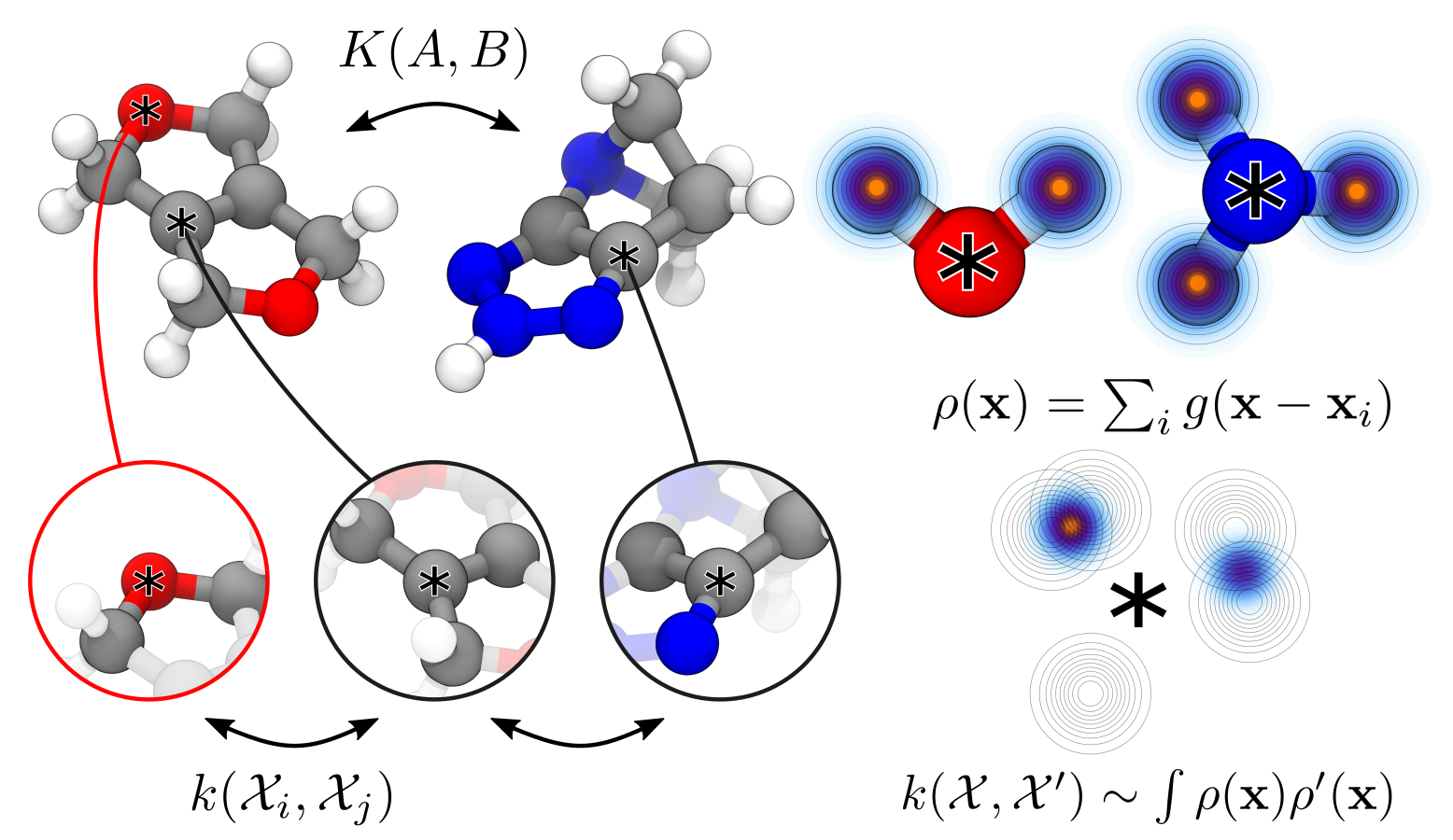}
\caption{A kernel function to compare solids and
molecules can be built based on density overlap kernels 
between atom-centered environments. {Chemical variability is accounted for by building separate neighbor densities for each distinct element~(see Ref.~\citenum{de+16pccp} and the SI).}
\label{fig:kernels}}
\end{figure}
 
\section{Materials and Methods}

Gaussian process regression (GPR) is a Bayesian machine learning framework\cite{RasmussenBook} which is also
formally equivalent to another  machine  learning  method,
Kernel  Ridge  Regression (KRR).
Both are based on a kernel function $K(x,x')$ that acts as a similarity measure between inputs $x$ and $x'$. Data points close in the metric space induced by the kernel are expected to correspond to 
the values $y$ and $y'$ of the function one is trying to approximate.
Given a set of training structures $x_i$ and the associated properties $y_i$, the prediction of the property for a new structure $x$ 
can be written as 
\begin{equation}
\bar y(x) = \sum_i w_i K(x, x_i),
\label{eq:krr}
\end{equation}
which is  a linear fit using the kernel function as a basis, evaluated at the locations of the prior observations. The optimal setting of the weight vector is ${\bf w} = ({\bf K}+\sigma^2{\bf I})^{-1} {\bf y}$, where $\sigma$ 
is the Tichonov regularization parameter.
In the framework of GPR,
which takes as its prior a multivariate normal distribution with the kernel as its covariance, Eq.~(\ref{eq:krr}) 
represents the mean, $\bar y$, of the posterior distribution
\begin{equation}
p(y^* | {\bf y}) \propto p(y^*\ \&\ {\bf y}) = {\cal N}(\bar y, \sigma^{*2})
\end{equation}
which now also provides an estimate of the error of the prediction, $\sigma^*$. 
The regularization parameter $\sigma$ corresponds to the expected deviation of the
observations from the underlying model due to statistical or systematic errors. 
Within GPR it is also easy to obtain generalizations for observations that are not of the function values, but linear functionals thereof (sums, derivatives). Low-rank (sparse) approximations of the kernel matrix are straightforward and help reduce the computational burden of the matrix inversion in computing the weight vector\cite{candela+rasmussen}.

The efficacy of machine learning methods critically depends on developing an appropriate kernel, or equivalently, on identifying relevant features in the input space that are used to compare data items. In the context of materials modelling, the input space of all possible molecules and solids is vast. We can drastically reduce the learning task by focusing on {\em local atomic environments} instead, and using a kernel between local environments as a building block. 

We use the Smooth Overlap of Atomic Positions (SOAP) kernel, which is the overlap integral of the neighbor density within a finite cutoff $r_c$, smoothed by a Gaussian with a length scale governed by the interatomic spacing, and finally integrated over all 3D rotations and normalized. This kernel is equivalent to the scalar product of the spherical power spectra of the neighbor density\cite{bart+13prb}, which
therefore constitutes a chemical {\em descriptor} of the neighbor environment. Both the kernel and 
the descriptor respect all physical symmetries (rotations, translations, permutations), are smooth functions of atomic coordinates and can be refined at will to provide a \emph{complete} description of each
environment.

To construct a kernel $K$ between two molecules (or periodic  structures) A and B from the SOAP kernel $k$ we average  over all possible pairs of environments,
\begin{equation}
K(A,B)=\sum_{i\in A, j\in B} P_{ij} \, k\left(x_i,x_j\right).
\label{eq:mol2env}
\end{equation} 
As shown in the SI, choosing $P_{ij} = \frac{1}{N_A N_B}$ for fitting the energy per atom is equivalent to defining it as a sum of atomic energy contributions (i.e. an interatomic potential), with the atomic energy function being a GPR fit using the SOAP kernel as its basis. 
Given that the available  observations 
are total energies and their derivatives with respect to atoms (forces), the learning machine provides us with the optimal decomposition of the quantum mechanical total energy into atomic contributions. In keeping with the nomenclature of the recent literature, we call a GPR model of the atomistic potential energy surface a ``Gaussian Approximation Potential'' (GAP), and a ``SOAP-GAP model'' is one which uses the SOAP kernel.

Other choices of $P$  are possible and will make sense for various applications. For example, setting $P$ to be the permutation matrix that maximizes the value of $K$ corresponds to the ``best match'' assignment between constituent atoms in the two structures that are compared - which can be computed in polynomial time by formulating the task as an optimal assignment problem~\cite{kuhn55nrlq}. It is possible to smoothly interpolate between the %
average 
and best match kernels using an entropy-regularized Wasserstein distance~\cite{cutu13nips} construction.

\section{Acknowledgments}
ABP was supported by a Leverhulme Early Career Fellowship and the Isaac Newton Trust until 2016, ABP also acknowledges support from CCP-NC funded by EPSRC (EP/M022501/1).
SD was supported by the NCCR MARVEL, funded by the Swiss National Science Foundation. MC acknowledges funding
by the European Research Council under the European Union’s Horizon 2020 research and innovation programme (grant agreement no. 677013-HBMAP).
CP and JRK were supported by the EU grant ``NOMAD'', grant no. 676580.
JRK acknowledges support from the EPSRC under grants EP/L014742/1 and EP/P002188/1. GC acknowledges support from EPSRC grants EP/L014742/1, EP/J010847/1 and EP/J022012/1. The work of NB was funded by the Office of Naval Research through the U. S. Naval Research Laboratory’s core basic research program.
 Computations were performed at the Argonne Leadership Computing Facility  under contract DE-AC02-06CH11357, the High Performance Computing Service at Cambridge University, computing resources provided by STFC Scientific Computing Department's SCARF cluster, and also ARCHER under the ``UKCP'' EPSRC grants EP/K013564/1 and EP/P022561/1.

\emph{Data Availability}
All data needed to evaluate the conclusions in the paper are present in the paper, in the cited references and/or the Supplementary Materials. Additional data related to this paper may be requested from the authors.

\emph{Competing Interests:} The authors declare that they have no competing interests.

\emph{Author contributions:} APB, SD, GC and MC performed and analysed calculations on molecular databases. CP, GC and MC performed and analysed drug binding predictions. APB, NB, JRK and GC performed and analysed calculations on silicon surfaces. All the authors contributed to the writing of the manuscript.

\nocite{bart+10prl,WGAP,aCgap,Perdew:1992jd,CASTEP,VanDuin:2001ud,Buehler:2006cm,Lenosky:2000wy,Tersoff:1988to,STILLINGER:1985vxa,solares2005,Sadowski:1994uj,OBoyle:2011cw,Wang:2004fe,Stewart:2012fa,MOPAC2016:online,Gaussian09:program,Ramakrishnan:2014ij,Becke:1993is,STEPHENS:1994jt,Curtiss:1998jm,MOLPRO_brief,fabe+17jctc,huo2017unified,Krishnan:1980kt,ceri+13jctc,bart+13prb,mont+13njp,de+16pccp,de+16pccp,ropo+16sd,mari+16jctc,soapxx}

\bibliographystyle{Science}

\foreach \x in {1,2,3,4,5,6,7,8,9}
{%
\clearpage


\section*{Supplementary material}

# Machine Learning Unifies the Modelling of Materials and Molecules
## Supporting Materials

Albert P. Bartók,[1] Sandip De,[2,3] Carl Poelking,[4] Noam Bernstein,[5] James Kermode,[6] Gábor Csányi,[7] and Michele Ceriotti[3]

[1] *Scientific Computing Department, Science and Technology Facilities Council, Rutherford Appleton Laboratory, Oxfordshire OX11 0QX, United Kingdom*
[2] *National Center for Computational Design and Discovery of Novel Materials (MARVEL)*
[3] *Laboratory of Computational Science and Modelling, Institute of Materials, Ecole Polytechnique Fédérale de Lausanne, Lausanne, Switzerland*
[4] *University Chemical Laboratory, University of Cambridge*
[5] *Center for Materials Physics and Technology, U.S. Naval Research Laboratory, Washington, DC 20375, USA*
[6] *Warwick Centre for Predictive Modelling, School of Engineering, University of Warwick, Coventry CV4 7AL, United Kingdom*
[7] *Engineering Laboratory, University of Cambridge, Trumpington Street, Cambridge CB2 1PZ, United Kingdom*

### section 1. The atom-centered GAP is equivalent to the average molecular kernel

Consider the KRR expression for the average energy per atom of a molecule $A$:

$$E(A)/N_A = \sum_n w_n K(A, A_n), \quad (1)$$

where $K(A, A')$ is a kernel function that measures the similarity between the molecule $A$ and a set of reference molecules $\{A_n\}$. The weights $w_n$ can be optimized by requiring that, for each molecule in such reference set, the energy predicted by Eqn. (1) matches that evaluated with an explicit quantum calculation, $E_n$. A similar expression can be written for an atom-based energy decomposition

$$\mathcal{E}(\mathcal{X}) = \sum_i \omega_i k(\mathcal{X}, \mathcal{X}_i), \quad (2)$$

with the difference that now the kernel function measures the similarity between *atomic environments* $\mathcal{X}$, and that the KRR evaluates the contribution to the total energy originating from an individual atom. This atom-based decomposition is the conventional way to define an interatomic potential, and has previously been used to create GAPs for materials [6, 44, 45].

To see how these two expressions are related to each other, consider that in a Gaussian process regression framework the kernel between two molecules is the same as the covariance between their energies $\langle E(A)E(B)\rangle = N_A N_B K(A,B)$. Similarly, the kernel between atomic environments, is the covariance between the atomic energies, $\langle \mathcal{E}(\mathcal{X}_i)\mathcal{E}(\mathcal{X}_j)\rangle = k(\mathcal{X}_j, \mathcal{X}_i)$. Under the assumptions that the energy decomposition is fully additive, so that $E(A) = \sum_{i \in A} \mathcal{E}(\mathcal{X}_i)$, one can see that

$$K(A,B) = \frac{\langle E(A)E(B)\rangle}{N_A N_B} = \\ = \frac{1}{N_A N_B} \sum_{i \in A, j \in B} k(\mathcal{X}_i, \mathcal{X}_j). \quad (3)$$

By substituting this expression for $K(A,B)$ into Eq. 1, it is possible to transform the expression of $E(A)$ into a sum over atom-based energies as in Eq. 2. Learning molecular energies using "structure" kernels that are equal to averages of atoms-centered kernels is thus equivalent to learning an atom-based energy decomposition using kernels between atomic environments.

### section 2. A SOAP-GAP potential for silicon

The configurations comprising the training set of the SOAP-GAP model for silicon are summarized in Table **S1**. The structures were generated by DFT molecular dynamics, starting from an initial structure of the given type, and using loose convergence settings of the DFT parameters. After collecting decorrelated samples, the energies, forces and virials were recalculated with tighter convergence settings of the parameters: 250 eV plane wave cutoff and a k-point density of 0.03 Å$^{-1}$. The PW91[46] exchange-correlation functional was used throughout. All calculations were carried with with the CASTEP package[47]. The crack tip structures were generated using an earlier GAP model that did not include those configurations. The structures with low coordination (with sp and sp$^2$ hybridizations) were included because it was found that without training on them the GAP model had a tendency to predict too low energies for such structures. Although not fully automated, this is motivated by the active learning approach, and the idea that to fully capture a probability distribution (here the Boltzmann distribution corresponding to the potential) it is not enough to specify where the probability is high (low energy structures) but also where it is low (the high energy structures).

We do not optimise the hyperparameters of the Gaussian process, because previous experience shows that—consistent with the Bayesian approach—our physically motivated guesses are good enough. With the database size required for the desired accuracy, the dependence of

| Structure type | # atoms | # structures | # inducing points | $\sigma_{\text{energy}}$ | $\sigma_{\text{force}}$ | $\sigma_{\text{virial}}$ |
|---|---:|---:|---:|---:|---:|---:|
| isolated atom | 1 | 1 | 1 | 0.001 | - | - |
| diamond | 2<br>16<br>54<br>128 | 104<br>220<br>110<br>55 | 500 | 0.001 | 0.1 | 0.05 |
| beta-tin | 2<br>16<br>54<br>128 | 60<br>220<br>110<br>55 | 500 | 0.001 | 0.1 | 0.05 |
| hexagonal | 1<br>8<br>27<br>64 | 110<br>30<br>30<br>53 | 500 | 0.001 | 0.1 | 0.05 |
| liquid | 64<br>128 | 69<br>7 | 500 | 0.003 | 0.15 | 0.2 |
| amorphous | 64<br>216 | 31<br>128 | 1000 | 0.01 | 0.2 | 0.4 |
| diamond surface (001) | 144 | 29 | 500 | 0.001 | 0.1 | 0.05 |
| diamond surface (110) | 108 | 26 | 500 | 0.001 | 0.1 | 0.05 |
| diamond surface (111)<br>   unreconstructed<br>   adatom<br>   Pandey<br>   DAS 3x3 unrelaxed | <br>96<br>146<br>96<br>52 | <br>47<br>11<br>50<br>1 | <br>500<br>250<br>500<br>100 | 0.001 | 0.1 | 0.05 |
| diamond vacancy | 63<br>215 | 100<br>111 | 500 | 0.001 | 0.1 | 0.05 |
| diamond divacancy | 214 | 78 | 500 | 0.001 | 0.1 | 0.05 |
| diamond interstitial | 217 | 115 | 500 | 0.001 | 0.1 | 0.05 |
| small (110) crack tip | 200 | 7 | 500 | 0.001 | 0.1 | 0.05 |
| small (111) crack tip | 192 | 10 | 500 | 0.001 | 0.1 | 0.05 |
| screw dislocation core | 144 | 19 | 200 | 0.001 | 0.1 | 0.05 |
| $sp^2$ bonded | 8 | 51 | 200 | 0.001 | 0.1 | 0.05 |
| sp bonded | 4 | 100 | 200 | 0.01 | 0.2 | 0.4 |
| Total | 169455 | 2148 | 8451 | | | |

**table S1**. **Summary of the database for the silicon model.** The total number of atoms corresponds to the entire database. The fitted potential has the unique label `GAP_2017_5_20_60_4_23_20_512`

the fit to the hyperparameters are quite weak. The locality of silicon is defined by the decay of the density matrix, and prior calculations indicate that force errors below 0.1 eV/Å are achievable with a cutoff of around 5 Å. The width parameter of the Gaussian functions that make up the neighbour density was 0.5 Å, close to the atomic unit of 1 Bohr, which is the typical length scale over which the potential energy varies. The truncation of the spherical harmonic expansion is a tradeoff between computational efficiency and accuracy - we typically fit a potential using tight tolerances, and as a last step, reduce the number of basis components as much as possible without compromising the accuracy. The regularisation parameters in the Gaussian process correspond to the expected accuracy, and are determined by the above locality criterion for the forces, and the estimated errors in the total energy and the virial due to the finite k-point sampling in the DFT calculations. Some high energy configurations (e.g. liquid and sp-bonded) have larger regularisation parameters.

The model is a sum of two terms. In addition to the SOAP-GAP term, we used a simple pair potential, parametrised to reproduce the dissociation and close-range repulsion behaviour of the Si dimer. The main purpose of the pair model is to augment the GPR model at short bond distances, where the energy scale is much larger compared to the attraction of interatomic bonding.

The options to the GAP fitting program to generate the SOAP-GAP model were

```
at_file=all_data.xyz gap={soap l_max=12 n_max
   =10 atom_sigma=0.5 zeta=4 cutoff=5.0
   cutoff_transition_width=1.0 central_weight
   =1.0 config_type_n_sparse={divacancy:500:
   interstitial:500:crack_110_1-10:500:
   surface_111:500:surface_110:500:sp2:200:sp
   :200:crack_111_1-10:500:dia:500:
   isolated_atom:1:bt:500:screw_disloc:200:sh
   :500:liq:500:surface_001:500:amorph:1000:
   surface_111_pandey:500:vacancy:500:111
   adatom:250:surface_111_3x3_das:100} delta
   =3.0 f0=0.0 covariance_type=dot_product
   sparse_method=cur_points} default_sigma
   ={0.001 0.1 0.05 0.0} config_type_sigma={
   liq:0.003:0.15:0.2:0.0:amorph
   :0.01:0.2:0.4:0.0:sp:0.01:0.2:0.4:0.0}
   energy_parameter_name=dft_energy
   force_parameter_name=dft_force
   virial_parameter_name=dft_virial
   config_type_parameter_name=config_type
   sparse_jitter=1.0e-8 e0_offset=2.0
   core_param_file=glue.xml core_ip_args={IP
   Glue}
```

The other interatomic potentials shown in the main paper for the DAS reconsutrctions were ReaxFF[48, 49], a Modified Embedded Atom Model[50], a Tersoff model[51] and the Stillinger-Weber model[52]. The data for the DFT curve was obtained from Solares et al.[53], which we

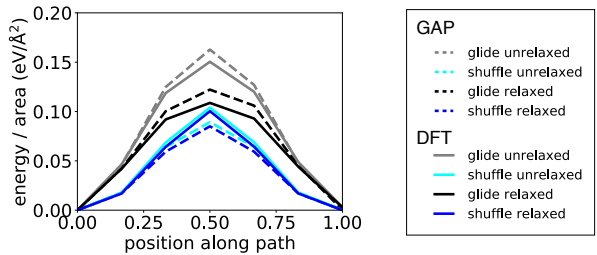

figure **S1**. **Stacking fault energetics for the silicon model.** The curves represent the energetics of paths that correspond to the formation of stacking faults in the diamond structure.

shifted by a constant to match the energy of the (111) unreconstructed surface calculated using our DFT parameters and setup.

The detailed analysis of the accuracy of the GAP model in comparison to other widely used potentials will be published elsewhere. Previously published works by some of us as well as other groups indicate that nonparametric fits such as GAP are capable of reproducing with good accuracy the energetics of a wide variety of configurations that are close to those present in their training set. Our results go beyond this by (i) showing exquisite accuracy in subtle situations such as the surface reconstructions shown in the main text (including extrapolation to large system sizes), and (ii) good transferability to configurations very far from those in the training set and also away from local minima. A demonstration of the latter is in Fig. **S1**, which shows the energetics of paths the correspond to the formation of two kinds of stacking faults. The highest error is less than 15% for the glide set case, and much lower for the shuffle set - other potentials typically have 30%-50% error.

**section 3. Predicting atomization energies for the GDB9 and QM7b databases**

**A. Computational details**

DFT geometries and energies were obtained from the original GDB9 database[21]. To generate the PM7-optimised geometries, we started with the SMILES strings in the GDB9. We used the CORINA program (version 3.60 0066)[54] to construct three-dimensional models of the molecules and to obtain initial Cartesian coordinates. A small fraction of the molecules failed to convert, for these we used OpenBabel[55] (version 2.3.0). As a part of the conversion, hydrogen atoms were added to the structures by CORINA and OpenBabel, then the configurations were relaxed by CORINA's built-in force field and the GAFF force field[56], respectively. The resulting configurations were further relaxed at the PM7 level of semi-empirical model[57] using the MOPAC

3

program[58] (versions 16.043L and 17.048L).

We adopted the relaxed geometries in the GDB9 database[60], and we carried out geometry relaxations on the oligopeptides using the Gaussian 09 program[59]. To maintain consistence with the GDB9 database, we used the same level of theory (Density Functional Theory and the B3LYP functional[61, 62]) and the 6-31G(2df,p) basis set[63]. CCSD(T) energetics of the DFT-relaxed configurations were calculated with MOLPRO[64] (version 2012.1), using the 6-311G** basis set[67].

Unless otherwise stated, in this section we discuss learning DFT energies based on DFT-optimized geometries. While this is largely an academic exercise, given that in order to obtain DFT structures one inevitably must compute DFT energies, it has often been used as a benchmark and so it is well-suited to make our error analysis directly comparable with previous studies. Results for learning CCSD(T) energies based on DFT geometries follow very similar trends, while learning based on PM7-optimized geometries presents a different sets of challenges that are discussed in the main text.

### B. Training set selection and error distribution

The GDB9 dataset contains - by construction - a relatively uneven sampling of chemical compound space, with some stoichiometries more heavily represented than others. A random selection of reference structures would give more thorough sampling of the densely populated regions, which might be advantageous to reduce the average error, but would leave extended portions of chemical space completely off the chart. An alternative approach would aim for a uniform sampling, so as to cover the margins of the distribution as well as the densely sampled regions. Farthest-point sampling (FPS) provides a simple, greedy algorithm to achieve this goal: given a set $\mathbb{S}_m = \{A_{i=1...m}\}$ of molecules selected out of the overall database $\mathbb{D}$, the next molecule to be included is determined by

$$A_{m+1} = \mathrm{argmax}_{A \in \mathbb{D}} \left[ \min_{A' \in \mathbb{S}} D(A, A') \right], \qquad (4)$$

where $D$ is the kernel-induced metric $D(A,B)^2 = K(A,A) + K(B,B) - 2K(A,B)$. Intermediate sampling methods, that balance diversity and relevance of the chosen molecules, are also possible [68]. Since Eqn. (4) relies solely on structural information, it is a practical strategy to decide where to invest computational resources to obtain a comprehensive sampling of the relevant chemical space. Fixing a maximum acceptable value of the minimum distance to the existing references, this approach also naturally extends to active learning. Whenever a new structure encountered in a simulation based on a ML potential is farther from the training set than this threshold, its energy can be computed with a high-end quantum calculation and the model be retrained on the extended reference set.

As shown in Figure **S2**, the strategy to select training points has a significant impact on the distribution of errors. Even though a FPS selection leads to a marginal increase of the MAE relative to a randomized choice, it enables a significant reduction of RMS. When studying the convergence of machine-learning methods, one should not stop at the MAE but also consider higher norms, that contain more information on the outliers, and the worst-case scenarios.

### C. Training curves and hyperparameter optimization

The SOAP kernel contains several adjustable parameters – that determine its completeness, evaluation cost, and the scale of interactions [39]. The parameters to the `glosim.py` package used to generate the kernel matrix for the production calculations were

```
~/source/glosim/glosim.py datafile.xyz -n 9 -l
    9 -g 0.3 -c 3 --zeta 2 --periodic --nonorm
    --kernel average
```

The code is available from http://cosmo-epfl.github.io. Internally, `glosim.py` called the SOAP routines in quippy, using the template

```
"soap central_reference_all_species=F
    central_weight=1.0 covariance_sigma0=0.0
    atom_sigma="+str(g)+" cutoff="+str(c)+"
    cutoff_transition_width=0.5 n_max="+str(n)
    +" l_max="+str(l)
```

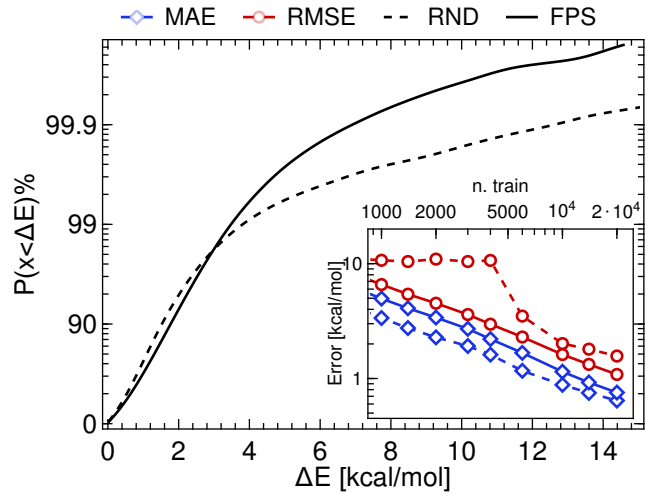

figure S2. **Error distribution for the GDB9 training.** Fraction of test configurations with a error smaller than a given threshold, for $n_{\mathrm{train}} = 20,000$ training structures selected at random (dashed line) or by FPS (full line). The inset shows the learning curves resulting from the two selection strategies, comparing the mean absolute error (blue) and root mean-square error (red).

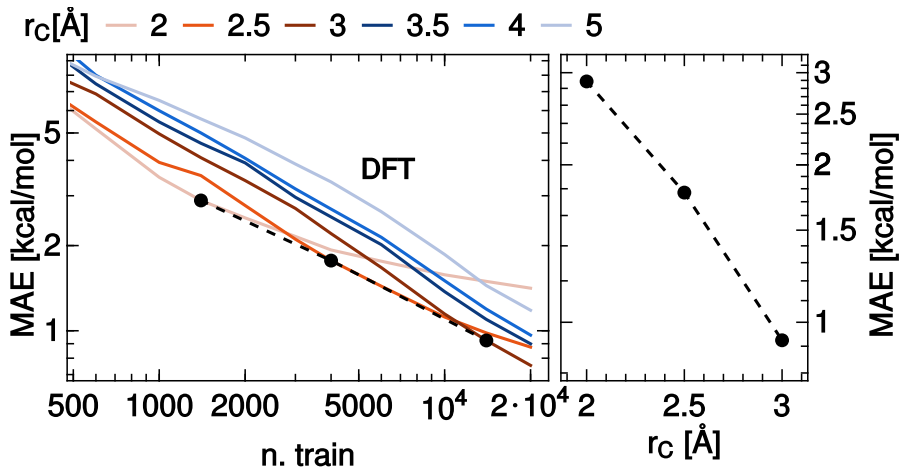

**figure S3**. **Optimal range of interactions for learning GDB9 DFT energies.** (left) Learning curves for the GDB9 dataset. 20,000 structures were selected by FPS and used for training DFT energetics, using the DFT geometries as inputs. Tests were performed on the remaining 114,000 structures. Different curves correspond to varying cutoffs in the SOAP environment selection, resulting in a different trend in the curve. Shorter cutoffs typically give smaller errors with small $n_{\text{train}}$, but the error saturates for larger train set size. The dashed curve highlights the envelope of the various training curves, signifying which value of $r_C$ gives the best performance for each training set size. The same points, plotted as a function of $r_C$ (right) give a sense of the energy scale of the interactions that can be modelled with local description over the specified range.

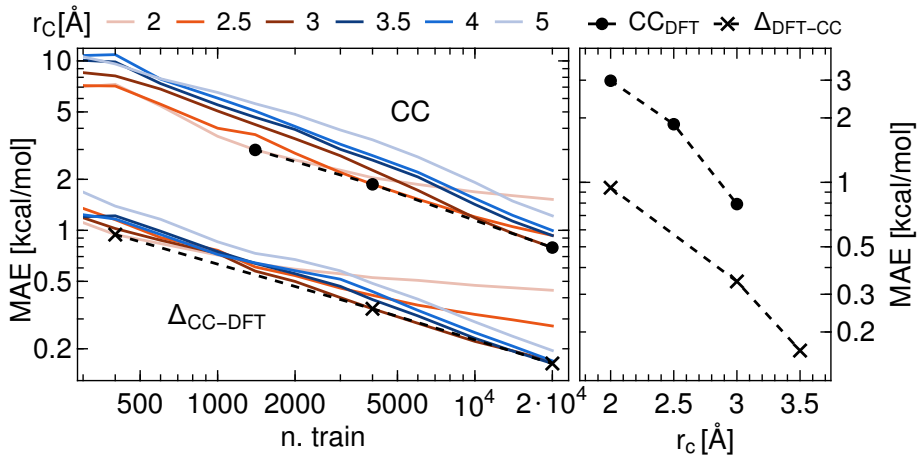

**figure S4**. **Optimal range of interactions for learning GDB9 CC and $\Delta_{\text{CC-DFT}}$ energies.** (left) Learning curves for the GDB9 dataset. 20,000 structures were selected by FPS and used for training CC energetics, using the DFT geometries as inputs. The top curves correspond to the error resulting from learning the full CC energy, whereas the lower set of curves correspond to the error that can be achieved using DFT energies as baseline. Tests were performed on 17,000 randomly-selected GDB9 structures, excluding those that were part of the train set. The dashed black curves highlight the envelope of the various training curves, signifying which value of $r_C$ gives the best performance for each training set size. The same points, plotted as a function of $r_C$ (right) give a sense of the energy scale of the interactions that can be modelled with local description over the specified range.

We did not optimize systematically the parameter space, but focused on the cutoff radius $r_C$ that enters the definition of local environments. As shown in Figures **S3** and **S4**, this exercise does not only make it possible to optimize the test error for a given size of the training set, but reveals information on the energy scale associated with different degrees of locality. DFT and CC energies both seem to exhibit a similar trend, with an energy scale of the order of 3 kcal/mol for a very short-range cutoff $r_C = 2$ Å, that decreases below 1 kcal/mol with $r_C = 3$ Å. When considering $\Delta_{\text{CC-DFT}}$, instead, the absolute energy scale is much lower, and one sees that longer-range interactions are crucial: in order to reach an accuracy below 0.2 kcal/mol, $r_C = 3.5$ Å is the best choice for the SOAP environment cutoff.

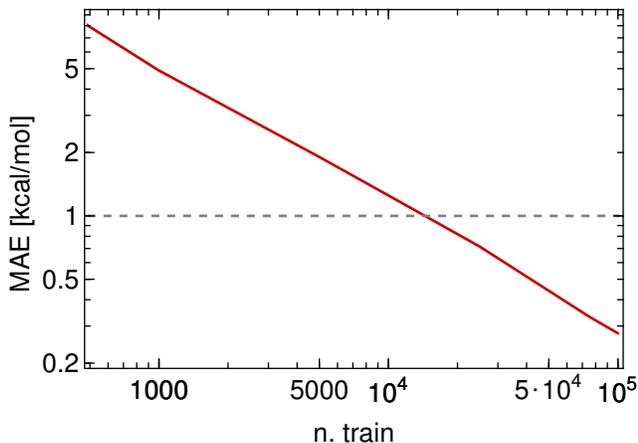

**figure S5**. **Training curves for the prediction of DFT energies using DFT geometries as inputs for the GDB9 dataset**. We selected about 33k structures as random to be used as a test set, and then sorted the remaining 100k structures in FPS order, and computed the MAE as a function of the number of inputs included in the training. We used the same kernel parameters as in the main text, and only increased the cutoff distance to 3.5 Å, to be able to capture finer-grained energetics. The figure demonstrates that the SOAP-GAP model is far from having reached its limiting accuracy when using 20k training inputs. For $n_{\text{train}} = 100,000$ the MAE drops below 0.28 kcal/mol.

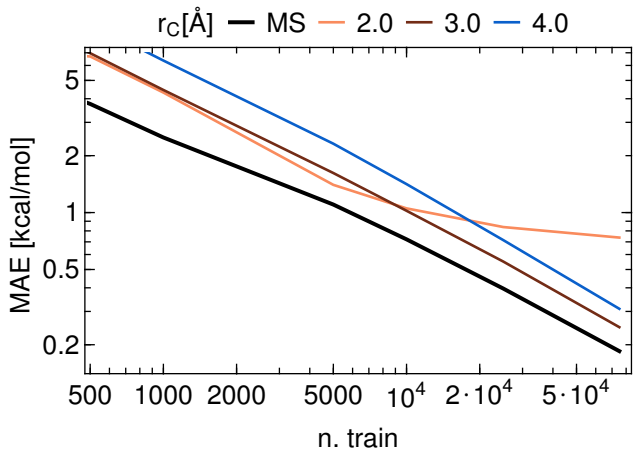

**figure S7**. **Training curves for the prediction of DFT energies using DFT geometries as inputs for the GDB9 dataset.** The error is computed on 34,000 randomly-selected structures, the training is performed on structures selected in FPS order from the remaining 100,000 configurations. The training curves for different SOAP cutoff length follow a similar trend to what observed for CC energies, and for QM7b. The thicker black curve, labelled MS (for multi-scale) uses a compound kernel built by averaging together the three kernels with different cutoff lengths.

### D. DFT-on-DFT benchmarks

Although in this work we focused on obtaining *useful* predictions, that would allow one to circumvent expensive electronic-structure calculations, most of the benchmarks in recent literature have been performed using DFT-optimized geometries as the input for predicting DFT energetics. In order to compare with other state-of-the art machine-learning models, and to provide an idea of the limiting accuracy and the scope for improvement for the SOAP-GAP model, we have also performed this kind of benchmark calculations. Figure **S5** demonstrates the behavior of the SOAP-GAP model for the GDB9 database when the number of training points is increased above 20,000. One can see that the error is far from saturating, and a MAE below 0.3 kcal/mol can be achieved with the same simple class of kernels we used in the main text, by increasing the training set to contain 100k structures.

We also attempted a preliminary demonstration of the possible directions in which one could improve the performance of SOAP-GAP kernels for a fixed training set size. For these tests we used the smaller QM7b dataset [8], that contains 7,211 molecules with up to 7 N,O,C,Cl,S atoms, with different degrees of H saturation. Our early study applying SOAP descriptors to this system Ref. [20], where we used a considerably more complex non-additive kernel with far from optimal parameter settings, demonstrated 1 kcal/mol MAE with 75% of the data set used for training. With the same training-set size, the present, much simpler, additive SOAP-GAP framework achieves

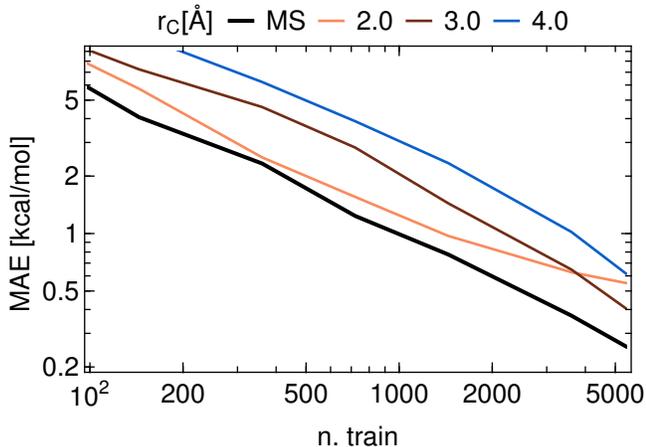

**figure S6**. **Training curves for the prediction of DFT energies using DFT geometries as inputs for the QM7b dataset.** Structures are selected in FPS order, and the error is computed on the remainder of the 7,211 configurations. The training curves for different SOAP cutoff length follow a similar trend to what is observed for the GDB9, with a trade-off between completeness of the description, and the extrapolative power for small training set size. The thicker black curve, labelled MS (for multi-scale) uses a compound kernel built by averaging together the three kernels with different cutoff lengths.

a MAE of 0.4 kcal/mol with a cutoff of 3 Å. The dependence of the training curves on cutoff radius is similar to what we observed for the GDB9 (Figure **S6**), with a tradeoff between the ultimate attainable accuracy and the extrapolative power for small training set size.

A very simple approach to improve the accuracy of our framework even further entails combining information from different length scales. Within a Bayesian formalism, one can just build a linear combination of different kernels, weighted by a factor that represents the relative contribution to the target property. Such a multi-scale kernel (specifically, one built as $k_{\text{MS}} = (256 k_{r_c=2} + 16 k_{r_c=3} + 1 k_{r_c=4})/273$) reduces the MAE consistently across training set sizes, reaching a MAE of just 0.26 kcal/mol with a training-set containing 75% of the overall data (Fig. **S6**). The same combination of kernels also enables dramatic improvements in the prediction of DFT energies for GDB9. As shown in Fig. **S7** using a multi-scale kernel combining information from 2, 3, 4 Å makes it possible to reach MAE below 1 kcal/mol with about 5,000 training points, that drops to a minuscule 0.18 kcal/mol by the time the train set contains 75,000 structures. Both the results on GDB9 and on QM7b are considerably better than similar benchmark calculations on these two databases [65, 66].

Another direction in which the SOAP descriptors can be improved involves using a choice other than $\kappa_{\alpha\beta} = \delta_{\alpha\beta}$ in the "alchemical" component of the kernel. $\kappa_{\alpha\beta}$ represents the "overlap" between different elements in the definition of the SOAP kernel, that is

$$k(\mathcal{X}, \mathcal{X}') = \int dR \left| \sum_{\alpha\beta} \kappa_{\alpha\beta} \int d\mathbf{x} \rho_\alpha(\mathbf{x}) \rho'_\beta(R\mathbf{x}) \right|^2, \quad (5)$$

where $\rho_\alpha$ and $\rho'_\beta$ correspond to the densities stemming from the species $\alpha$ and $\beta$ in the environments $\mathcal{X}$ and $\mathcal{X}'$ respectively (see Ref. [20] for a more thorough discussion). We did not attempt a systematic study of the role of these hyperparameters – that represent correlations between the properties of different elements – but experimented with a definition of the form $\kappa_{\alpha\beta} = e^{-(a_\alpha - a_\beta)^2/2\Delta^2}$, where $a_\alpha$ represent an atomic property. Results are promising: using the first ionization energy for $a$ and $\Delta = 1$ eV we obtained (for DFT-on-DFT QM7b, with the reference 75% training set size, and a SOAP cutoff of 3 Å) a MAE of 0.38 kcal/mol. Using the electron affinity and $\Delta = 1$ eV, we obtained a MAE of 0.34 kcal/mol. Using Pauling electronegativity and $\Delta = 0.5$ we achieved a MAE of 0.33 kcal/mol.

### E. Oligopeptides

To test the extrapolation capabilities of the SOAP-GAP model built on the GDB9, we considered a few hundred structures from a database of gas-phase conformers of proteinogenic oligopeptides [25]. We picked in

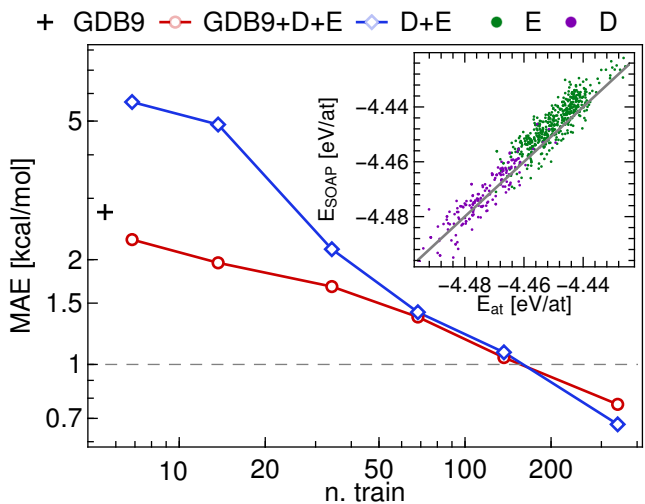

**figure S8**. **Errors in the learning of conformational stability of dipeptides based on GDB9.** Training curves for the prediction of DFT energies using DFT geometries as inputs, for a dataset containing a total of 684 configurations of glutamic acid dipeptide (E) and aspartic acid dipeptide (D). The inset shows the correlation between DFT and ML energies as obtained from the model trained on 20,000 FPS-selected structures from the GDB9, which has a MAE of 2.8 kcal/mol – and would already be sufficient for a preliminary screening of candidate conformers. The model can be systematically enhanced by including FPS-selected conformations from the oligopeptide dataset. With about 20% of the structures, both the extendend GDB9 model and a model trained directly on the oligopeptides conformers reaches the 1 kcal/mol milestone.

particular 500 local minima for glutammic acid dipeptide (E) and for 184 local minima for aspartic acid dipeptide (D) (containing respectively 14 and 13 non-H atoms), and re-optimized the geometries using exactly the same density-functional protocol as used for the GDB9. We then proceeded to test the performance of the GDB9-trained models in predicting the relative stability of the different conformers. We started from the rather academic exercise of using DFT-optimized geometries to predict DFT energetics. As shown in Fig. **S8**, GDB9-trained model provides predictions with an accuracy comparable to DFT – not only of the absolute stability of the two compounds, but also of the relative stability of different conformers. The model can be improved systematically by including structures from the oligopeptides dataset.

Figure **S9** shows an analysis of the predictive power of the GDB9-trained model for the DFT-to-CC corrections $\Delta_{\text{DFT-CC}}$. Not only can the SOAP-GAP model correct the large discrepancy between the DFT and the CC atomization energies for the two compounds – which can be largely ascribed to atomic corrections, but it can also provide some degree of correction to the *relative energetics* of different conformers of the two molecules – which is remarkable when one considers that this kind of data

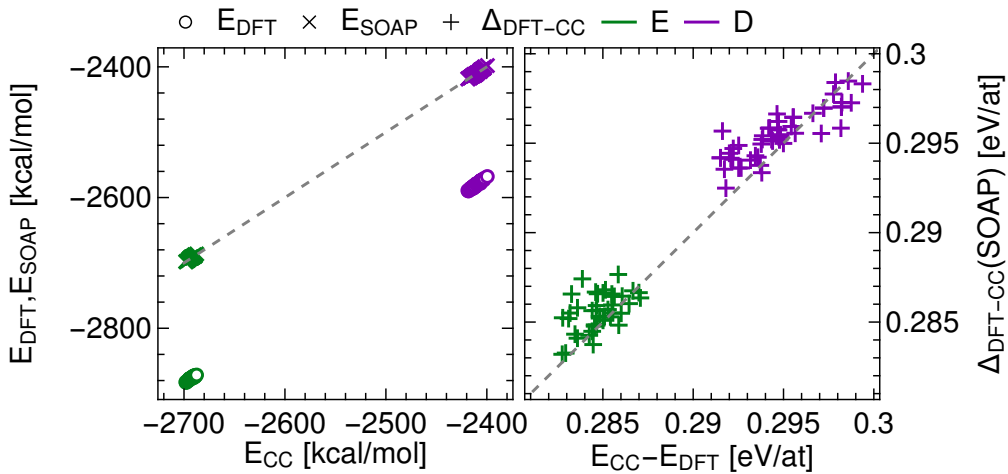

figure S9. Correlation plots for the learning of the energetics of dipeptide configurations, based on GDB9. (left) Correlation between DFT and CC atomization energies for 41 conformers of glutammic acid dipeptide (E) and 52 conformers of aspartic acid dipeptide (D). Disks correspond to the actual DFT and CC energies, crosses correspond to DFT energies corrected with the $\Delta_{\text{DFT-CC}}$ term obtained by the GDB9-trained model. (right) Correlation between the actual difference $E_{\text{CC}} - E_{\text{DFT}}$, and the model prediction.

is not explicitly included in the GDB9.

### F. Glucose

As shown in Fig. **S8**, when focusing on a restricted set of compounds, it can be sufficient to use just a handful of training configurations to obtain energy predictions on par with the most accurate electronic structure methods. We considered a set of 208 conformers of glucose, including both closed and open-chain configurations [26]. We use the same SOAP kernel parameters as for the GDB9, and train the model on 20 structures, selected by FPS, using the remaining 188 for validation. As discussed in the main text, this brings the typical error in the energy of conformers relative to benchmark, complete-basis-set CCSD(T) values to less than 0.2-0.4 kcal/mol when using DFT as a baseline, corresponding to a reduction between 50 and 80% of the MAE, relative to the intrinsic discrepancy between the two methods.

### section 4. Ligand Classification and Visualisation

The classification of ligands from the DUD-E into actives and inactives was performed with a Kernel-Support-Vector-Machine with a 1-norm penalty factor $C = 1.0$. The decision function for a test structure $B$ is then

$$z_B = \sum_A \alpha_A^* y_A K(A, B) + \beta^*, \quad (6)$$

where $y_A \in {-1, +1}$ is the class label of a structure $A$ from the training set. The predicted class for $B$ is $\hat{y}_B = \text{sign}(z_B)$. $\beta^*$ determines the decision threshold, and the coefficients $\alpha_A^*$ are computed based on the optimisation problem (in its dual formulation):

Maximize

$$\sum_A \alpha_A - \frac{1}{2} \sum_{A,A'} y_A \alpha_A K(A, A') y_{A'} \alpha_{A'}, \quad (7)$$

subject to

$$\sum_A y_A \alpha_A = 0, \quad 0 \leq \alpha_A \leq C. \quad (8)$$

The kernel $K(A, B)$ is chosen as either an average-kernel or "best-match" SOAP (MATCH, in practice a REMatch kernel with $\gamma = 0.01$ [20]). The training is performed on sets of compounds comprising the same number of actives and inactives (decoys), thus automatically assigning equal weight to both classes. SOAP descriptors were generated with `soapxx` software[69] and the following parameters.

```
"soap-atom": {
"spectrum.global": false,
"spectrum.gradients": false,
"spectrum.2l1_norm": false,
"radialbasis.type" : "gaussian",
"radialbasis.mode" : "adaptive",
"radialbasis.N" : 9,
"radialbasis.sigma": 0.5,
"radialcutoff.Rc": 3.5,
"radialcutoff.Rc_width": 0.5,
"radialcutoff.type": "heaviside",
"radialcutoff.center_weight": 1.0,
"angularbasis.type": "spherical-harmonic",
"angularbasis.L": 6,
"exclude_centers": [],
"exclude_targets": [],
```

```
"type_list": ["Br", "C", "Cl", "F", "H", "I",
    "N", "O", "P", "S"]
}
```

For MATCH, the contribution $\delta z_{j,B}$ of an individual atomic environment $j \in B$ to $z_B$ was computed by decomposing the decision function via the permutation matrix $P_{ij}$:

$$\delta z_{j,B} = \sum_A \alpha_A^* y_A \sum_{i \in A} P_{ij} k_{ij}(A, B) + \frac{\beta^*}{\|B\|}. \quad (9)$$

Here, $k_{ij}(A, B)$ is the SOAP kernel between atomic environments $i \in A$ and $j \in B$. We visualised the atomic contributions by defining a density ("binding field") $\rho_B(\boldsymbol{r}) = \sum_{j \in B} \delta z_{j,B} \mathcal{N}(\boldsymbol{r}_j, \sigma_j)$, made up of atom-centered Gaussians $\mathcal{N}$ of width $\sigma_j = 0.5$ Å. This density is subsequently visualised on an isosurface of the atomic density on which the SOAP descriptor is built.

All the ligand binding predictions and binding field maps are available at http://www.libatoms.org/dude-soap and individual PDFs for each ligand can be downloaded from http://www.libatoms.org/dude-soap/pdf/.


}

\end{document}